\documentclass[letterpaper,letter,twocolumn]{./jpsj3}

\usepackage{graphicx} 
\usepackage{color} 
\usepackage{amsmath}
\usepackage{amsfonts}
\usepackage{amssymb}

\usepackage{bm}
\usepackage{txfonts}

\newcommand{\subm}[1]{_{\mathrm {#1}}}

\newcommand{\etal}{\textit{et~al.}}
\newcommand{\degc}{$^{\circ}$C}
\renewcommand{\deg}{^{\circ}}

\newcommand{\Tcz}{\ensuremath{T\subm{c0}}}

\newcommand{\Hcc}{\ensuremath{H\subm{c2}}}

\newcommand{\chin}{\ensuremath{\chi\subm{n}}}

\newcommand{\chis}{\chi\subm{sc}}

\newcommand{\sro}{Sr$_2$RuO$_4$}

\newcommand{\Hp}{\ensuremath{H\subm{P}}}
\newcommand{\C}{c}

\newcommand{\corrected}[1]{#1}

\title{Specific-heat evidence of the first-order superconducting transition in Sr$_{\bm{2}}$RuO$_{\bm{4}}$}

\author{%
Shingo~Yonezawa\thanks{E-mail address: yonezawa@scphys.kyoto-u.ac.jp},
Tomohiro~Kajikawa, Yoshiteru~Maeno
}

\inst{%
Department of Physics, Graduate School of Science, Kyoto University, Kyoto 606-8502, Japan
}

\abst{
We investigate the specific heat of ultra-pure single crystals of Sr$_2$RuO$_4$, a leading candidate of a spin-triplet superconductor.
We for the first time obtained specific-heat evidence of the first-order superconducting transition below 0.8~K, namely divergent-like peaks and clear hysteresis in the specific heat at the upper critical field.
The first-order transition occurs for all in-plane field directions.
The specific-heat features for the first-order transition are found to be highly sensitive to sample quality; in particular, the hysteresis becomes totally absent in a sample with slightly lower quality.
These thermodynamic observations provide crucial bases to understand the unconventional pair-breaking effect responsible for the first-order transition.
}

\kword{%
Sr$_2$RuO$_4$, spin-triplet superconductivity, first-order transition, specific heat, superconducting phase diagram, pair-breaking effects
}

\recdate{\today}

\begin{document}

\maketitle

Breaking mechanisms of superconductivity by the magnetic field is of fundamental interest, because pair-breaking mechanisms, directly related to interactions between superconductivity and magnetic field, provide rich information about the superconducting (SC) state.
In case of type-II superconductors, the ordinary pair-breaking is the orbital effect, which originates from the orbital motion of Cooper pairs circulating around quantized vortices~\cite{TinkhamText}.
When the orbital effect dominates, the SC to normal state (S-N) transition at the upper critical field $\Hcc$ is a second-order transition (SOT).
The orbital effect is also characterized by the linear temperature dependence of $\Hcc$ near the zero-field critical temperature $\Tcz$.
Another mechanism, which is less common, is the Pauli effect, which originates from the competition between the Zeeman spin polarization and the SC condensation energy~\cite{Clogston1962}. 
The Pauli effect is characterized by a typical concave-down shape of the $\Hcc(T)$ curve and often causes a first-order transition (FOT) at $\Hcc$ at low temperatures~\cite{Matsuda2007JPhysSocJpnReview}.
Such features for the Pauli effect has been observed in several spin-singlet superconductors such as heavy fermions~\cite{Ikeda2001,Bianchi2002,Aoki2007.JPhysSocJpn.76.063701,Kasahara2007.PhysRevLett.99.116402}, organics~\cite{Tanatar2002PhysRevB,Lortz2007PhysRevLett,Beyer2012.PhysRevLett.109.027003,Yonezawa2008.PhysRevLett.100.117002,Yonezawa2012.PhysRevB.85.140502R}, and iron pnictides~\cite{Burger2013.PhysRevB.88.014517,Zocco2013.PhysRevLett.111.057007,Kittaka2013.JPhysSocJpn.83.013704}.
In some special cases, exotic pair-breaking effects can emerge. 
For example, in the ferromagnetic superconductor UCoGe, a new type of pair breaking has been suggested where the magnetic field directly weakens the pairing ``glue''~\cite{Hattori2012.PhysRevLett.108.0664403}.

Recently, possible unconventional pair-breaking mechanism in \sro\ has been suggested by the present authors~\cite{Yonezawa2013.PhysRevLett.110.077003}.
This oxide, with $\Tcz \sim 1.5$~K, has been extensively studied for 20 years as a leading candidate for spin-triplet superconductors~\cite{Maeno1994,Mackenzie2003RMP,Maeno2012.JPhysSocJpn.81.011009,Kallin2012.RepProgPhys.75.042501}.
The spin-triplet state has been directly revealed by comprehensive spin susceptibility $\chi\subm{spin}$ measurements by the nuclear magnetic resonance (NMR) using several atomic sites~\cite{Ishida1998.Nature.396.658,Ishida2001.PhysRevB.63.060507R,Murakawa2004.PhysRevLett.93.167004} and the polarized neutron scattering~\cite{Duffy2000.PhysRevLett.85.5412}:
Both kinds of  experiments have revealed that $\chi\subm{spin}$ in the SC state $\chis$ is equal to that in the normal state $\chin$.
In addition, other unconventional SC phenomena attributable to the orbital and/or spin degrees of freedom in the SC wave function support spin-triplet scenarios~\cite{Luke1998.Nature.394.558,Nelson2004.Science.306.1151,Xia2006.PhysRevLett.97.167002,Kashiwaya2011.PhysRevLett.107.077003,Nakamura2011.PhysRevB.84.060512R,Jang2011.Science.331.186,Anwar2013.SciRep.3.2480}.
More recently, the spin susceptibility is measured again using Ru and Sr NMR, and the spin-triplet nature is further confirmed~\cite{Ishida2014.unpublished,Miyake2014.JPhysSocJpn.83.053701}.

Interestingly, we have recently revealed by means of the magnetocaloric-effect (MCE) study that the S-N transition of \sro\ is of first order when the magnetic field is parallel to the $ab$ plane and when the temperature is below 0.8~K~\cite{Yonezawa2013.PhysRevLett.110.077003}.
Clearly, the FOT contradict with the ordinary orbital effect, which leads to a SOT.
Existence of a pair-breaking mechanism different from the orbital effect for $H\parallel ab$ is strongly supported by the recent small-angle neutron scattering  study~\cite{Rastovski2013.PhysRevLett.111.087003}.
\corrected{In addition, for \sro, the Pauli effect is inconsistent with the experimental result of $\chis \simeq \chin$ by NMR~\cite{Ishida1998.Nature.396.658,Ishida2001.PhysRevB.63.060507R,Murakawa2004.PhysRevLett.93.167004,Ishida2014.unpublished} and polarized neutron scattering~\cite{Duffy2000.PhysRevLett.85.5412}, because the characteristic pair-breaking field for the Pauli effect, the Pauli limit $\Hp$, should be infinitely large since $\Hp\propto (\chin-\chis)^{-1/2}$.}
Therefore, a new pair-breaking mechanism responsible for the FOT should exist.

In order to obtain new insights into the origin of the FOT, we measured the specific heat $\C$ of ultra-pure single crystals of 
\sro. 
Importantly, $\C/T$, being equal to the temperature derivative of the entropy $S$, is directly related to the quasiparticle density of states and is seldom affected by vortex pinning. 
Thus, $\C/T$ is a fundamental physical quantity in the study of superconductors. 
In contrast, the MCE is essentially a ``magnetic'' measurement since one obtains $\partial S/\partial H = \partial M/\partial T$ from the MCE, where $M$ is the magnetization. 
This magnetic nature is clearly demonstrated by the fact that the MCE can be affected by vortex pinning, which makes the above thermodynamic relation sometimes invalid.
In this Letter, we report the first specific-heat evidence of the FOT, namely divergent-like peak and hysteresis in $\C(H)/T$ at $\Hcc$.
We also found that the FOT features in specific heat is quite sensitive to the sample quality.
The observation of the FOT in a fundamental quantity should provide an important bases toward resolution of the origin of the unusual pair-breaking effect. 

For the present study, we used single crystals of \sro\ grown by the floating-zone method~\cite{Mao2000.MaterResBull.35.1813}.
In this Letter, we mainly report results on Sample~\#2 ($\Tcz=1.50$~K, 0.184~mg), although some results on Sample~\#1 ($\Tcz=1.45$~K, 0.684~mg) are also shown for comparison.
These samples are identical to those used in the MCE experiment~\cite{Yonezawa2013.PhysRevLett.110.077003}.
The value of $\Tcz$ was obtained by ac susceptibility and specific heat measurements in zero field.
We emphasize that $\Tcz$ of Sample~\#2 is equal to the ideal $\Tcz$ of \sro\ in the clean limit~\cite{Mackenzie1998.PhysRevLett.80.161},
indicating its extreme cleanness.
The specific heat $\C$ was measured using a conventional ac method~\cite{Sullivan1968}.
We chose 0.8~Hz (Sample~\#1) and 1.8-2.4~Hz (Sample~\#2) as the frequency of the temperature modulation for the sample heat capacity measurement.
We confirmed absence of frequency dependence in the obtained heat capacity near this frequency range, indicating that the measurement has been performed with proper frequencies. 
The background heat capacity was also measured and we found no detectable field strength and angle dependences in the present field range.
In this Letter, we present $\varDelta\C(H)/T\equiv [C(H)-C(H>\Hcc)]/nT$. Here, $C$ is the measured heat capacity including the background and $n$ is the molar amount the sample. Note that the subtraction of $C(H>\Hcc)$ allows us to subtract the normal-state specific heat (including the phonon contribution) as well as the background heat capacity. 
The magnetic field was applied using a vector magnet system~\cite{Deguchi2004RSI}.
Throughout this Letter, we denote $\theta$ as the polar angle of the field with respect to the $c$ axis, and $\phi$ as the azimuthal angle within the $ab$ plane with respect to the $a$ axis.
The accuracy in $\theta$ is better than $0.1\deg$ for both samples; while the accuracy in $\phi$ is around $1\deg$ for Sample~\#2 and $5$-$10\deg$ for Sample~\#1.


\begin{figure}
\begin{center}
\includegraphics[width=8.5cm]{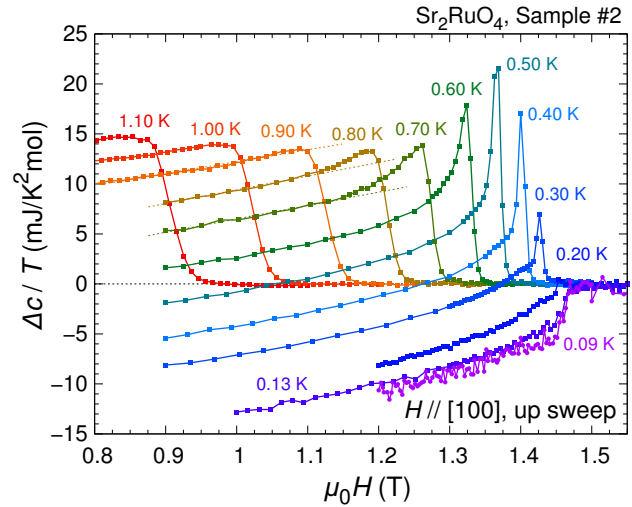}
\end{center}
\caption{
(Color online)
Field dependence of $\varDelta\C/T$ of \sro\ (Sample~\#2) for $H\parallel [100]$. Only field-up sweep data are shown in these figures.
The broken lines present results of linear fitting to $\varDelta\C/T$ data at 0.7, 0.8 and 0.9~K. 
Clearly, divergent-like feature exists below 0.8~K but not above 0.9~K.
\label{fig:C-H_100}
}
\end{figure}

In Fig.~\ref{fig:C-H_100}(a), we present $\varDelta \C(H)/T$ for $H \parallel [100]$ with various temperatures. 
At high temperatures, the step-like anomaly in $\C(H)$ at $\Hcc$ indicates the SOT, as expected for ordinary type-II superconductors. 
However, as temperature is decreased, the anomaly at $\Hcc$ starts to acquire a divergent-like peak features.
This peak in $\C/T$, indicating a discontinuous change in $S$ at $\Hcc$, evidences the FOT.
The divergent-like feature becomes most pronounced at around 0.5~K.
This behavior is consistent with simple thermodynamics for a FOT, as being discussed latter.
By comparing the data near $\Hcc$ with the linear extrapolation of the low-field data (Fig.~\ref{fig:C-H_100}), 
we can see that the peak at $\Hcc$ starts to appear below 0.8~K but absent above 0.9~K.
Thus we estimate the temperature below which the FOT occurs, $T\subm{FOT}$, as $\sim 0.8$~K. 
These results are consistent with the previous MCE results~\cite{Yonezawa2013.PhysRevLett.110.077003}.

\begin{figure}
\begin{center}
\includegraphics[width=8.5cm]{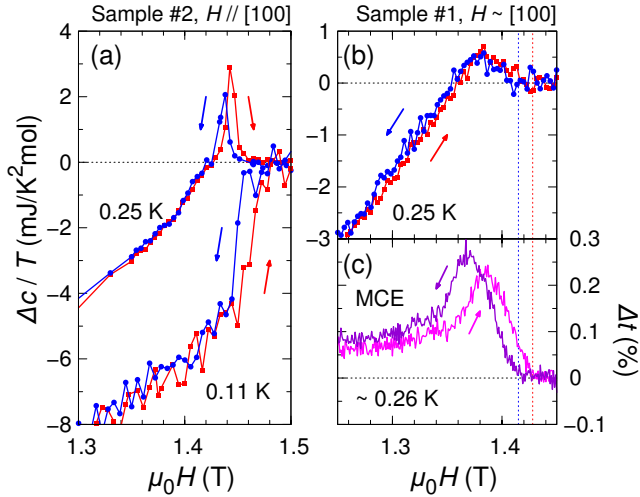}
\end{center}
\caption{
(Color online)
Comparison between field-up (red squares) and down sweep (blue circles) data of $\varDelta\C(H)/T$ (a) of Sample~\#2 for $H\parallel [100]$ and (b) of Sample~\#1 for $H\sim [100]$. 
(c) The relative temperature change $\varDelta t$ due to the MCE~\cite{Yonezawa2013.PhysRevLett.110.077003} for Sample~\#1 at $T\sim 0.26$~K measured with the sweep rate of 1.02~mT/sec.
The vertical broken lines indicate $\Hcc$ deduced from the field-up and down sweep data of the MCE.
\label{fig:C-H_hysteresis}
}
\end{figure}

As further definitive evidence for the FOT, we succeeded in observing hysteresis in $\C(H)$.
As presented in Fig.~\ref{fig:C-H_hysteresis}(a), we observed a clear difference between the onset $\Hcc$ of field-up and down sweeps.
At 0.11~K, the difference between $\Hcc$ of the up- and down-sweeps is approximately 20~mT, \corrected{in agreement with} the MCE result~\cite{Yonezawa2013.PhysRevLett.110.077003}.
To the best of our knowledge, except for the MCE, hysteresis at $\Hcc$ of \sro\ has never been reported. 
Thus, the present result marks the first observation of hysteresis with a ``point-by-point'' measurement technique, with each data point taken at a fixed $H$.
Note that, generally speaking, it is more difficult to observe hysteresis with a point-by-point measurement, compared with continuous measurement techniques such as the MCE: 
The supercooled/superheated metastable state can exhibit transition to the stable state within the time required for one data point, which is rather long for the former kind of techniques. 
\corrected{It should be noted here that the ac method for specific-heat measurements sometimes hinder hystereses, in particular, for {\em temperature} sweeps. Indeed, we could not observe any hystereses in {\em temperature} sweeps even for Sample~\#2 (not shown). Nevertheless, the clear hystereses observed in {\em field} sweeps, with the hysteresis width in agreement with the MCE, indicate that the use of the ac method is appropriate in the present case.}

Interestingly, the FOT features in $\C(H)$ are found to be extremely sensitive to the sample quality.
Indeed, in Sample~\#1, with a slightly lower sample quality, 
we did not observe any sizable hysteresis in $\C(H)$ as exemplified in Fig.~\ref{fig:C-H_hysteresis}(b), although in the MCE measurement a clear hysteresis is seen even for this sample (Fig.~\ref{fig:C-H_hysteresis}(c)).
The results are naturally understood that the metastable supercooled/superheated state at the FOT can become easily unstable by existence of nucleation centers such as impurities or surface defects and is only detectable by a ``fast'' measurement such as the MCE.
This situation is very similar to water, where supercooled liquid water below 0\degc\ can be stable only if the water itself and the container are both clean enough.
Otherwise, the supercooled liquid water, if exists, changes into ice within a very short period of time.
Note that the peak of $\C(H)$ just below $\Hcc$ is quite vague in this sample.
This indicates that the FOT is very easily broadened by sample inhomogeneity, as commonly observed in other FOTs. 
The comparison between the two samples clearly indicates that the appearance of the FOT features strongly depends on sample quality, and that the extreme cleanness of Sample~\#2 allows us to observe hysteresis even with a ``point-by-point'' technique for the first time. 
This extreme sensitivity of the FOT features to the sample quality also explain why the hysteresis has not been observed in previous studies~\cite{NishiZaki2000JPhysSocJpn,Deguchi2002,Deguchi2004.PhysRevLett.92.047002,Deguchi2004.JPhysSocJpn.73.1313,Tenya2006.JPhysSocJpn.75.023702}.

\begin{figure}
\begin{center}
\includegraphics[width=8.5cm]{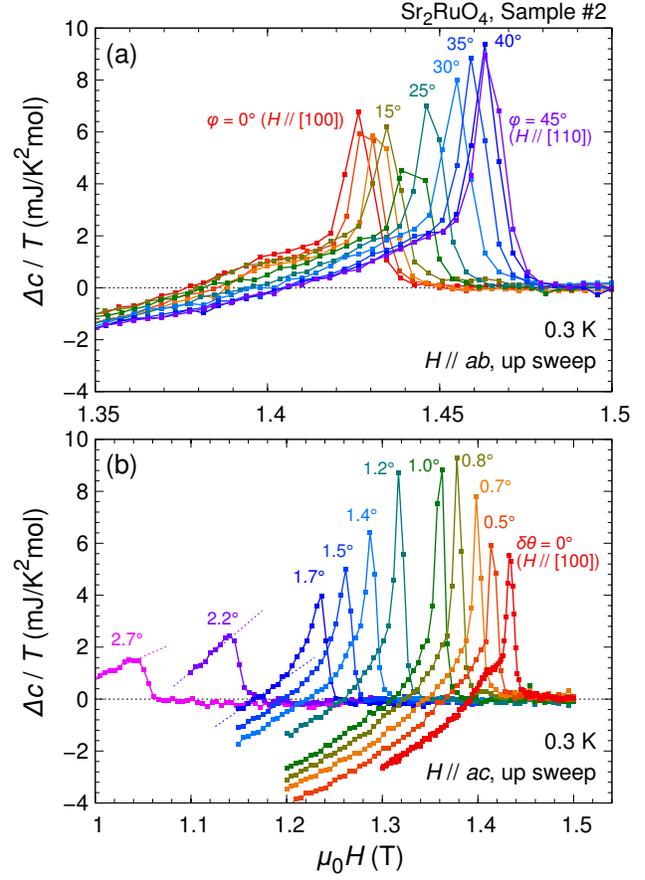}
\end{center}
\caption{
(Color online)
Field-strength dependence of $\varDelta \C/T$ at 0.3~K with different magnetic field directions.
(a) In-plane field-angle variation of $\varDelta \C(H)/T$ for $\phi=0$, 5, 10, \ldots, $45\deg$.
(b) Out-of-plane field-angle variation of $\varDelta \C(H)/T$ for fields in the $ac$ plane (i.e. $\phi=0\deg$).
The values near the curves indicate $\delta\theta \equiv 90\deg - \theta$, the field tilt angle with respect to the $ab$ plane.
The broken lines present results of linear fitting to $\varDelta\C/T$ data at $\delta\theta = 1.7$, 2.2, and $2.7\deg$.
\label{fig:C-H_phi-theta-dep}
}
\end{figure}

Next, we examine field-angle effects on the occurrence of the FOT.
In Fig.~\ref{fig:C-H_phi-theta-dep}(a), we present $\varDelta \C(H)/T$ at 0.3~K for several in-plane field directions.
The divergent-like peak is seen for all curves presented.
Therefore, it is now confirmed that the FOT occurs for all in-plane field directions.
Figure~\ref{fig:C-H_phi-theta-dep}(b) represents $\varDelta \C(H)/T$ at 0.3~K for fields slightly tilted away from the $ab$ plane.
Here we define the tilt angle $\delta\theta = 90\deg - \theta$.
As the field is tilted from the $ab$ plane, the divergent-like behavior, once weakly enhanced around $\delta\theta \sim 1.0\deg$, is rapidly suppressed for higher $\delta\theta$.
Although a small divergent-like feature remains at $\delta\theta = 1.7\deg$, it totally disappears at $\delta\theta = 2.2\deg$ and the specific-heat behavior at $\Hcc$ fully agrees with the SOT.
Thus, a change from the FOT to SOT at low temperature occurs around $\delta\theta \sim 2\deg$, being consistent with the previous MCE result~\cite{Yonezawa2013.PhysRevLett.110.077003}.
This change in the order of the S-N transition originates from the rapid recovery of the orbital-effect contribution caused by an increase of the out-of-plane field component.

We should comment here on the evolution of the shape of the $\varDelta\C(H)/T$ curves.
As temperature is decreased below $T\subm{FOT}$, the peak height grows toward 0.5~K but decrease again at lower temperatures.
This non-monotonic behavior can be consistently understood by using the Clausius-Clapeyron relation,
which manifests $\mu_0 \mathrm{d}\Hcc/\mathrm{d}T = -\varDelta S / \varDelta M$ at a FOT.
Here, $\varDelta S$ and $\varDelta M$ are the jumps in $S$ and $M$ at the FOT, respectively.
From the relation $(\partial S/\partial T)_H = \C/T$, the entropy jump can be obtained as $\varDelta S = \int (\varDelta\C/T) \mathrm{d}T$,
where the integral should be performed across the FOT. 
With a triangular assumption for the integral, i.e. $\int (\varDelta\C/T) \mathrm{d}T \sim (1/2)(\varDelta\C\subm{peak}/T)\varDelta T$, the peak height $\varDelta\C\subm{peak}/T$ is nearly proportional to $\varDelta S = -\mu_0 (\mathrm{d}\Hcc/\mathrm{d}T) \varDelta M$.
Because $\mathrm{d}\Hcc/\mathrm{d}T$ remains finite while $\varDelta M$ goes to zero as $T \to T\subm{FOT}$, the peak height decreases as $T\to T\subm{FOT}$.
In contrast, because $\mathrm{d}\Hcc/\mathrm{d}T$ decreases to zero while $\varDelta M$ remains finite as $T \to 0$, the peak height also decreases at low temperatures.
At lower temperatures, because $\varDelta\C/T$ is nearly equal to the entropy change with respect to the normal state~\cite{NoteLowTempC}, $\varDelta\C(H)/T$ is negative below $\Hcc$ (irrespective of the order of the S-N transition) and exhibits a sharp jump at the FOT.
Note that the jump in $\varDelta\C/T$ at 0.1~K is approximately $4.5\pm 0.5$~mJ/K$^2$mol (Figs.~\ref{fig:C-H_100}(a) and \ref{fig:C-H_hysteresis}(a)), which \corrected{is consistent with} the MCE result of the entropy jump $\varDelta S/T = 3.5 \pm 1$~mJ/K$^2$mol at 0.2~K~\cite{Yonezawa2013.PhysRevLett.110.077003}.
Similar argument can be applied also to the non-monotonic $\theta$ dependence of the peak in Fig.~\ref{fig:C-H_phi-theta-dep}(b).

\begin{figure}
\begin{center}
\includegraphics[width=8.5cm]{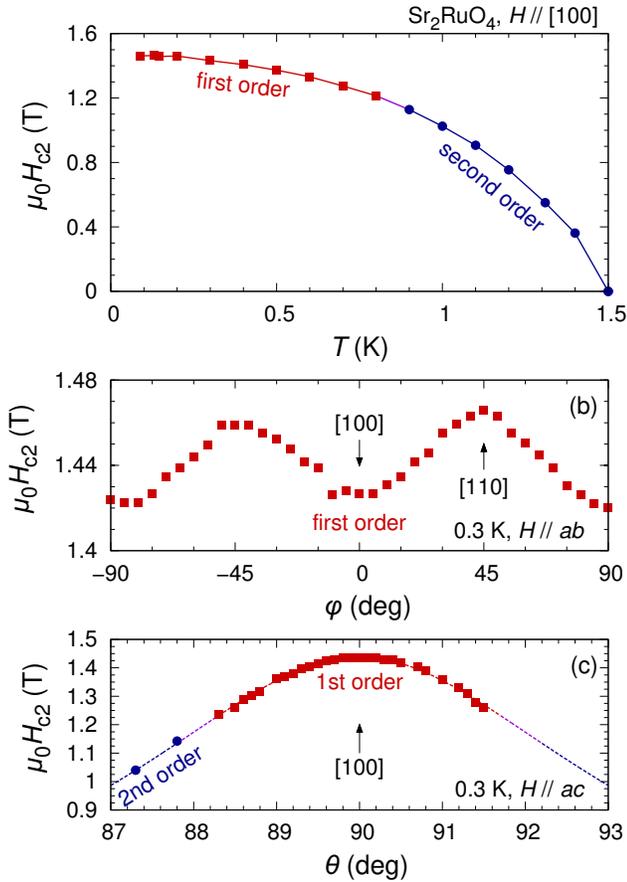}
\end{center}
\caption{
(Color online) 
(a) Superconducting $H$-$T$ phase diagram of \sro\ for $H\parallel [100]$.
(b) In-plane field angle $\phi$ dependence of $\Hcc$ at 0.3~K.
(c) Polar angle $\theta$ dependence of $\Hcc$ at 0.3~K.
For all panels, the red squares indicate $\Hcc$ with a FOT evidenced by divergent-like feature in $\delta\c(H)/T$ and the blue circles indicate $\Hcc$ with a SOT. The broken curve in (c) is a just guide to the eyes.
\label{fig:phase-diagrams}
}
\end{figure}

The present results are summarized in the phase diagrams presented in Fig.~\ref{fig:phase-diagrams}.
Here, we define $\Hcc$ as the peak position of $\varDelta\C(H)/T$ on the field-up sweep in the case of the FOT, and as the midpoint of the transition in the case of the SOT. 
As mentioned above, below $T\subm{FOT}\sim 0.8$~K, the S-N transition becomes first order.
Within the $ab$ plane, at 0.3~K, $\Hcc(\phi)$ exhibits four-fold sinusoidal oscillation with the maximum at $\phi=\pm 45\deg$ (i.e. $H\parallel [110]$), as shown in Fig.~\ref{fig:phase-diagrams}(b), being consistent with the previous studies~\cite{Mao2000.PhysRevLett.84.991,Deguchi2004.PhysRevLett.92.047002,Deguchi2004.JPhysSocJpn.73.1313,Kittaka2009.PhysRevB.80.174514}.
At this temperature, the S-N transition is of first order for all in-plane field directions.
As the magnetic field is tilted from the $ab$ plane, $\Hcc$ rapidly decreases and the S-N transition changes from the FOT to SOT around $\delta \theta = 2\deg$.

Below, we explain current situation toward the understanding of the origin of the unusual FOT in \sro.
As we explained before, the FOT cannot be attributed to the ordinary orbital effect. 
\corrected{Since the FOT at a first glance resembles the case of the Pauli effect, possibility of the Pauli effect has been recently examined again by Ishida \etal, who re-confirmed the absence of the spin-susceptibility decrease using Sr and Ru NMR, and even found a novel phenomenon attributable to the spin-triplet state~\cite{Ishida2014.unpublished,Miyake2014.JPhysSocJpn.83.053701}.}
To the best of our knowledge, alternative scenarios explaining the FOT in \sro\ have not been reported so far.
As proposed by Yanase~\cite{Yanase2014.JPhysSocJpn.83.061019}, the strong-coupling effect, or in other words, the feedback effect, should probably be taken into account to explain the FOT.
Recent angle-resolved photoemission spectroscopy (ARPES) with orbital or spin resolution~\cite{Iwasawa2010.PhysRevLett.105.226406,Veenstra2014.PhysRevLett.112.127002} reveals $k$-dependent spin and orbital locking, with pronounced effects particularly on the $\beta$ and $\gamma$ Fermi surfaces along the [110] direction, originating from the spin-orbit interaction of the order of $\lambda\subm{SOI} \sim 100$~meV.
Such spin-orbital locking in the underlying normal state may largely affect SC properties in \sro, although a microscopic theory predicts that the effect of the spin-orbit interaction to the SC state can be rather small due to a reduction by the factor $(\lambda\subm{SOI}/E\subm{F})^n$ ($n=1,\,2,\ldots$)~\cite{Yanase2014.JPhysSocJpn.83.061019}, where $E\subm{F}$ is the Fermi energy and is expected to be larger than $\lambda\subm{SOI}$.
It is now important not only to perform breakthrough experiments using ultra-clean single crystals or to build novel theories, but also to revisit previous experiments and theories with these new insights in mind, in order to clarify the origin of the FOT.

Before closing the discussion, we comment here that the clear multiple specific-heat anomaly reported by Deguchi \etal~\cite{Deguchi2002} was not reproduced in our present study. 
Although splitting into two peaks was commonly observed in a few samples in the previous study~\cite{DeguchiUnpublished}, these samples used in the previous study is 200 times larger than the sample used in the present study.
Thus, it cannot be excluded that the peak splitting was due to multiple crystal domains in the samples.
Nevertheless, another weak anomaly can be seen in the present data e.g.\ at 0.3~K \corrected{and $\sim 1.4$~T} for $H\parallel [100]$ (see Figs.~\ref{fig:C-H_100} and \ref{fig:C-H_phi-theta-dep}); at this point, we cannot exclude a possibility that this is due to either remaining tiny multiple domains or surface/edge contributions.

To summarize, we obtained evidence of the first-order SC transition of \sro\ by specific heat measurements, marking the first confirmation of the FOT with a point-by-point measurement technique. 
The appearance of the FOT features strongly depends on sample quality. 
Thus, in order to observe a clear FOT, it is crucially important to use an extremely clean single crystal. 
These results provide important bases toward resolution of the interesting puzzle on the origin of the FOT, namely the unknown pair-breaking mechanism in \sro.

\begin{acknowledgements}
We acknowledge Z. Q. Mao and F. H\"ubler for their contributions to crystal growth, and K.~Ishida, K.~Deguchi, K.~Machida, M.~Sigrist, Y.~Yanase, and T.~Nomura for useful discussions.
We also acknowledge KOA Corporation for providing us with their products for the calorimeter.
This work is supported by Grants-in-Aids for Scientific Research (KAKENHI 22103002, 23540407, 23110715, and 26287078) from MEXT and JSPS.
\end{acknowledgements}

\bibliography{../SRO-paper,%
D:/cygwin/home/Owner/SSP/paper/string,%
D:/cygwin/home/Owner/SSP/paper/TMTSF,%
D:/cygwin/home/Owner/SSP/paper/textbook,%
D:/cygwin/home/Owner/SSP/paper/FFLO,%
D:/cygwin/home/Owner/SSP/paper/CeCoIn5,%
D:/cygwin/home/Owner/SSP/paper/superconductors,%
D:/cygwin/home/Owner/SSP/paper/organic_SC,%
D:/cygwin/home/Owner/SSP/paper/SC,%
D:/cygwin/home/Owner/SSP/paper/Fe-Pn,%
D:/cygwin/home/Owner/SSP/paper/Sr2RuO4,%
D:/cygwin/home/Owner/SSP/paper/Sr3Ru2O7,%
D:/cygwin/home/Owner/SSP/paper/high-Tc,%
D:/cygwin/home/Owner/SSP/paper/heavy-Fermion,%
D:/cygwin/home/Owner/SSP/paper/measurement_technique%
}

\begin{thebibliography}{10}

\bibitem{TinkhamText}
M.~Tinkham: {\em Introduction to Superconductivity, Second Edition}
  (McGraw-Hill, New York, 1996).

\bibitem{Clogston1962}
A.~M. Clogston, Phys. Rev. Lett. {\bfseries 9}, 266 (1962).

\bibitem{Matsuda2007JPhysSocJpnReview}
Y.~Matsuda and H.~Shimahara, J. Phys. Soc. Jpn. {\bfseries 76}, 051005 (2007).

\bibitem{Ikeda2001}
S.~Ikeda, H.~Shishido, M.~Nakashima, R.~Settai, D.~Aoki, Y.~Haga, H.~Harima,
  Y.~Aoki, T.~Namiki, H.~Sato, and Y.~\=Onuki, J. Phys. Soc. Jpn {\bfseries
  70}, 2248 (2001).

\bibitem{Bianchi2002}
A.~Bianchi, R.~Movshovich, N.~Oeschler, P.~Gegenwart, F.~Steglich, J.~D.
  Thompson, P.~G. Pagliuso, and J.~L. Sarrao, Phys. Rev. Lett. {\bfseries 89},
  137002 (2002).

\bibitem{Aoki2007.JPhysSocJpn.76.063701}
D.~Aoki, Y.~Haga, T.~D. Matsuda, N.~Tateiwa, S.~Ikeda, Y.~Homma, H.~Sakai,
  Y.~Shiokawa, E.~Yamamoto, A.~Nakamura, R.~Settai, and Y.~{\={O}nuki}, J.
  Phys. Soc. Jpn. {\bfseries 76}, 063701 (2007).

\bibitem{Kasahara2007.PhysRevLett.99.116402}
Y.~Kasahara, T.~Iwasawa, H.~Shishido, T.~Shibauchi, K.~Behnia, Y.~Haga, T.~D.
  Matsuda, Y.~Onuki, M.~Sigrist, and Y.~Matsuda, Phys. Rev. Lett. {\bfseries
  99}, 116402 (2007).

\bibitem{Tanatar2002PhysRevB}
M.~A. Tanatar, T.~Ishiguro, H.~Tanaka, and H.~Kobayashi, Phys. Rev. B
  {\bfseries 66}, 134503 (2002).

\bibitem{Lortz2007PhysRevLett}
R.~Lortz, Y.~Wang, A.~Demuer, P.~H.~M. Bottger, B.~Bergk, G.~Zwicknagl,
  Y.~Nakazawa, and J.~Wosnitza, Phys. Rev. Lett. {\bfseries 99}, 187002 (2007).

\bibitem{Beyer2012.PhysRevLett.109.027003}
R.~Beyer, B.~Bergk, S.~Yasin, J.~A. Schlueter, and J.~Wosnitza, Phys. Rev.
  Lett. {\bfseries 109}, 027003 (2012).

\bibitem{Yonezawa2008.PhysRevLett.100.117002}
S.~Yonezawa, S.~Kusaba, Y.~Maeno, P.~Auban-Senzier, C.~Pasquier, K.~Bechgaard,
  and D.~J{\'{e}}rome, Phys. Rev. Lett. {\bfseries 100}, 117002 (2008).

\bibitem{Yonezawa2012.PhysRevB.85.140502R}
S.~Yonezawa, Y.~Maeno, K.~Bechgaard, and D.~J{\'{e}}rome, Phys. Rev. B
  {\bfseries 85}, 140502(R) (2012).

\bibitem{Burger2013.PhysRevB.88.014517}
P.~Burger, F.~Hardy, D.~Aoki, A.~E. Bohmer, R.~Eder, R.~Heid, T.~Wolf,
  P.~Schweiss, R.~Fromknecht, M.~J. Jackson, C.~Paulsen, and C.~Meingast, Phys.
  Rev. B {\bfseries 88}, 014517 (2013).

\bibitem{Zocco2013.PhysRevLett.111.057007}
D.~A. Zocco, K.~Grube, F.~Eilers, T.~Wolf, and H.~v.~L{\"{o}}hneysen, Phys.
  Rev. Lett. {\bfseries 111}, 057007 (2013).

\bibitem{Kittaka2013.JPhysSocJpn.83.013704}
S.~Kittaka, Y.~Aoki, N.~Kase, T.~Sakakibara, T.~Saito, H.~Fukazawa, Y.~Kohori,
  K.~Kihou, C.-H. Lee, A.~Iyo, H.~Eisaki, K.~Deguchi, N.~K. Sato, Y.~Tsutsumi,
  and K.~Machida, J. Phys. Soc. Jpn. {\bfseries 83}, 013704 (2013).

\bibitem{Hattori2012.PhysRevLett.108.0664403}
T.~Hattori, Y.~Ihara, Y.~Nakai, K.~Ishida, Y.~Tada, S.~Fujimoto, N.~Kawakami,
  E.~Osaki, K.~Deguchi, N.~K. Sato, and I.~Satoh, Phys. Rev. Lett. {\bfseries
  108}, 066403(1 (2012).

\bibitem{Yonezawa2013.PhysRevLett.110.077003}
S.~Yonezawa, T.~Kajikawa, and Y.~Maeno, Phys. Rev. Lett. {\bfseries 110},
  077003 (2013).

\bibitem{Maeno1994}
Y.~Maeno, H.~Hashimoto, K.~Yoshida, S.~Nishizaki, T.~Fujita, J.~G. Bednorz, and
  F.~Lichtenberg, Nature {\bfseries 372}, 532 (1994).

\bibitem{Mackenzie2003RMP}
A.~P. Mackenzie and Y.~Maeno, Rev. Mod. Phys. {\bfseries 75}, 657 (2003).

\bibitem{Maeno2012.JPhysSocJpn.81.011009}
Y.~Maeno, S.~Kittaka, T.~Nomura, S.~Yonezawa, and K.~Ishida, J. Phys. Soc. Jpn.
  {\bfseries 81}, 011009 (2012).

\bibitem{Kallin2012.RepProgPhys.75.042501}
C.~Kallin, Rep. Prog. Phys. {\bfseries 75}, 042501 (2012).

\bibitem{Ishida1998.Nature.396.658}
K.~Ishida, H.~Mukuda, Y.~Kitaoka, K.~Asayama, Z.~Q. Mao, Y.~Mori, and Y.~Maeno,
  Nature {\bfseries 396}, 658 (1998).

\bibitem{Ishida2001.PhysRevB.63.060507R}
K.~Ishida, H.~Mukuda, Y.~Kitaoka, Z.~Q. Mao, H.~Fukazawa, and Y.~Maeno, Phys.
  Rev. B {\bfseries 63}, 060507(R) (2001).

\bibitem{Murakawa2004.PhysRevLett.93.167004}
H.~Murakawa, K.~Ishida, K.~Kitagawa, Z.~Q. Mao, and Y.~Maeno, Phys. Rev. Lett.
  {\bfseries 93}, 167004 (2004).

\bibitem{Duffy2000.PhysRevLett.85.5412}
J.~A. Duffy, S.~M. Hayden, Y.~Maeno, Z.~Mao, J.~Kulda, and G.~J. McIntyre,
  Phys. Rev. Lett. {\bfseries 85}, 5412 (2000).

\bibitem{Luke1998.Nature.394.558}
G.~M. Luke, Y.~Fudamoto, K.~M. Kojima, M.~I. Larkin, J.~Merrin, B.~Nachumi,
  Y.~J. Uemura, Y.~Maeno, Z.~Q. Mao, Y.~Mori, H.~Nakamura, and M.~Sigrist,
  Nature {\bfseries 394}, 558 (1998).

\bibitem{Nelson2004.Science.306.1151}
K.~D. Nelson, Z.~Q. Mao, Y.~Maeno, and Y.~Liu, Science {\bfseries 306}, 1151
  (2004).

\bibitem{Xia2006.PhysRevLett.97.167002}
J.~Xia, Y.~Maeno, P.~T. Beyersdorf, M.~M. Fejer, and A.~Kapitulnik, Phys. Rev.
  Lett. {\bfseries 97}, 167002 (2006).

\bibitem{Kashiwaya2011.PhysRevLett.107.077003}
S.~Kashiwaya, H.~Kashiwaya, H.~Kambara, T.~Furuta, H.~Yaguchi, Y.~Tanaka, and
  Y.~Maeno, Phys. Rev. Lett. {\bfseries 107}, 077003 (2011).

\bibitem{Nakamura2011.PhysRevB.84.060512R}
T.~Nakamura, R.~Nakagawa, T.~Yamagishi, T.~Terashima, S.~Yonezawa, M.~Sigrist,
  and Y.~Maeno, Phys. Rev. B {\bfseries 84}, 060512(R) (2011).

\bibitem{Jang2011.Science.331.186}
J.~Jang, D.~G. Ferguson, V.~Vakaryuk, R.~Budakian, S.~B. Chung, P.~M. Goldbart,
  and Y.~Maeno, Science {\bfseries 331}, 186 (2011).

\bibitem{Anwar2013.SciRep.3.2480}
M.~S. Anwar, T.~Nakamura, S.~Yonezawa, M.~Yakabe, R.~I.~H. Takayanagi, and
  Y.~Maeno, Sci. Rep. {\bfseries 3}, 2480 (2013).

\bibitem{Ishida2014.unpublished}
{K}. Ishida {\textit{et al}}., \corrected{private communication}.

\bibitem{Miyake2014.JPhysSocJpn.83.053701}
K.~Miyake, J. Phys. Soc. Jpn. {\bfseries 83}, 053701 (2014).

\bibitem{Rastovski2013.PhysRevLett.111.087003}
C.~Rastovski, C.~D. Dewhurst, W.~J. Gannon, D.~C. Peets, H.~Takatsu, Y.~Maeno,
  M.~Ichioka, K.~Machida, and M.~R. Eskildsen, Phys. Rev. Lett. {\bfseries
  111}, 087003 (2013).

\bibitem{Mao2000.MaterResBull.35.1813}
Z.~Mao, Y.~Maeno, and H.~Fukazawa, Mater. Res. Bull. {\bfseries 35}, 1813
  (2000).

\bibitem{Mackenzie1998.PhysRevLett.80.161}
A.~P. Mackenzie, R.~K.~W. Haselwimmer, A.~W. Tyler, G.~G. Lonzarich, Y.~Mori,
  S.~Nishizaki, and Y.~Maeno, Phys. Rev. Lett. {\bfseries 80}, 161 (1998).

\bibitem{Sullivan1968}
P.~F. Sullivan and G.~Seidel, Phys. Rev. {\bfseries 173}, 679 (1968).

\bibitem{Deguchi2004RSI}
K.~Deguchi, T.~Ishiguro, and Y.~Maeno, Rev. Sci. Instrum. {\bfseries 75}, 1188
  (2004).

\bibitem{NishiZaki2000JPhysSocJpn}
S.~NishiZaki, Y.~Maeno, and Z.~Mao, J. Phys. Soc. Jpn. {\bfseries 69}, 572
  (2000).

\bibitem{Deguchi2002}
K.~Deguchi, M.~A. Tanatar, Z.~Mao, T.~Ishiguro, and Y.~Maeno, J. Phys. Soc.
  Jpn. {\bfseries 71}, 2839 (2002).

\bibitem{Deguchi2004.PhysRevLett.92.047002}
K.~Deguchi, Z.~Q. Mao, H.~Yaguchi, and Y.~Maeno, Phys. Rev. Lett. {\bfseries
  92}, 047002 (2004).

\bibitem{Deguchi2004.JPhysSocJpn.73.1313}
K.~Deguchi, Z.~Q. Mao, and Y.~Maeno, J. Phys. Soc. Jpn. {\bfseries 73}, 1313
  (2004).

\bibitem{Tenya2006.JPhysSocJpn.75.023702}
K.~Tenya, S.~Yasuda, M.~Yokoyama, H.~Amitsuka, K.~Deguchi, and Y.~Maeno, J.
  Phys. Soc. Jpn. {\bfseries 75}, 023702 (2006).

\bibitem{NoteLowTempC}
{This is easily obtained from the relation $\C/T = (\partial S/\partial T)_H
  \simeq [S(T)-S(0)]/[T-0] = S(T)/T$ for a small $T$, where $S(0)=0$ by
  definition.}

\bibitem{Mao2000.PhysRevLett.84.991}
Z.~Q. Mao, Y.~Maeno, S.~NishiZaki, T.~Akima, and T.~Ishiguro, Phys. Rev. Lett.
  {\bfseries 84}, 991 (2000).

\bibitem{Kittaka2009.PhysRevB.80.174514}
S.~Kittaka, T.~Nakamura, Y.~Aono, S.~Yonezawa, K.~Ishida, and Y.~Maeno, Phys.
  Rev. B {\bfseries 80}, 174514 (2009).

\bibitem{Yanase2014.JPhysSocJpn.83.061019}
Y.~Yanase, S.~Takamatsu, and M.~Udagawa, J. Phys. Soc. Jpn. {\bfseries 83},
  061019 (2014).

\bibitem{Iwasawa2010.PhysRevLett.105.226406}
H.~Iwasawa, Y.~Yoshida, I.~Hase, S.~Koikegami, H.~Hayashi, J.~Jiang,
  K.~Shimada, H.~Namatame, M.~Taniguchi, and Y.~Aiura, Phys. Rev. Lett.
  {\bfseries 105}, 226406 (2010).

\bibitem{Veenstra2014.PhysRevLett.112.127002}
C.~Veenstra, Z.-H. Zhu, M.~Raichle, B.~Ludbrook, A.~Nicolaou, B.~Slomski,
  G.~Landolt, S.~Kittaka, Y.~Maeno, J.~Dil, I.~Elfimov, M.~Haverkort, and
  A.~Damascelli, Phys. Rev. Lett. {\bfseries 112}, 127002 (2014).

\bibitem{DeguchiUnpublished}
{K}. Deguchi {\textit{et al}}., \corrected{private communication}.

\end{thebibliography}

\end{document}